\documentclass[amsmath,amssymb,aps,prl,twocolumn,showpacs,superscriptaddress]{revtex4-1} 

\usepackage{graphicx}

\begin{document}
\title{Paramagnetic tunneling systems and their contribution \\to the polarization echo in glasses \\ (extended)}
\author{A.~Borisenko}
\email{borisenko@kipt.kharkov.ua}
\affiliation{
NSC Kharkiv Institute of Physics and Technology, Akademichna 1,
61108 Kharkiv, Ukraine 
}
\affiliation{
Dipartimento di Fisica e Matematica, Universit\'a degli Studi dell'Insubria, via Valleggio 11, 22100 Como, Italy
}
\author{G.~Jug} 
\affiliation{
Dipartimento di Fisica e Matematica, Universit\'a degli Studi dell'Insubria, via Valleggio 11, 22100 Como, Italy
}
\affiliation{
INFN-Sezione di Pavia and CNR-IPCF Unit\'a di Roma, Italy
}
\date{\today}
\begin{abstract}
Startling magnetic effects on the spontaneous polarization echo in some silicate glasses at low and ultra-low
 temperatures have been reported in the last decade or so. Though some progress in search of an explanation 
has been made by considering the nuclear quadrupole dephasing of tunneling particles, here we show that the effect 
of a magnetic field can be understood quantitatively by means of a special tunnel mechanism associated with paramagnetic 
impurities. For the Fe$^{3+}$-, Cr$^{3+}$- and Nd$^{3+}$-contaminated glasses we provide reasonable fits to the published data as a function 
of applied magnetic field and temperature.
\end{abstract}

\maketitle

Anomalous low-temperature magnetic properties of some silicate glasses have attracted a considerable experimental and theoretical interest since the unexpected findings by Strehlow et al. \cite{Strehlow}. 
Since the low-temperature behavior of glasses is believed to be governed by  tunneling systems (TS) (see e.g. \cite{TM}), a lot of theoretical effort has 
been applied to account for the coupling of TS to a magnetic field. All the theoretical models that have been proposed so far may be grouped into either 
"orbital" \cite{Mexican_hat,Dimerization,Jug} or "spin" \cite{Quadrupole,PTS} type according to the assumed mechanism for the TS coupling to a magnetic field. 

A model of paramagnetic tunneling systems (PTS), belonging to the "spin" group, has been recently proposed, in collaboration, by one of the present authors  \cite{PTS}. 
It considers an effective charged particle with an intrinsic spin $S$, tunneling between the coordinate states (wells)  $\left| l \right\rangle$  and  $\left| r \right\rangle$  of a 1D double-well potential.
%
%
In each coordinate state the spin is quantized along the axis $z$ and the only spin projections allowed are  $S_{z} =\pm S$ . 
Therefore, one can treat both coordinate and spin states of such PTS in the spin-1/2 formalism. Let two sets of Pauli matrices 
$\hat{\boldsymbol{\sigma}}$  and  $\hat{\boldsymbol{\tau}}$ together with the corresponding $2\times2$ unity matrices $\hat{1}_{\sigma}$ 
and $\hat{1}_{\tau}$  refer to the PTS coordinate and spin states, respectively. Then the Hamiltonian of such PTS is as follows \cite{Borisenko}:
\begin{align}\label{eq1}
&\hat{H}_{\rm{PTS}} = \nonumber \\
& -\tfrac{1}{2} \left(h\hat{1}_{\tau}\otimes\hat{\sigma}_{z} +\delta \hat{1}_{\tau}\otimes\hat{\sigma}_{x} +u\hat{\tau}_{z}\otimes\hat{1}_{\sigma} -d\hat{\tau}_{y}\otimes \hat{\sigma}_{y} \right),
\end{align}
where $h$ is the energy difference (asymmetry) between the states  $\left| l \right\rangle$ and $\left| r \right\rangle$, $u$ is the energy difference between the states  $\left| \pm S \right\rangle$  
in each well and the tunneling matrix elements are $\delta =\Delta \cos \left(\alpha /2\right)$, $d=\Delta \sin \left(\alpha /2\right)$,
%
%
$\Delta$  being a tunneling matrix element between the states  $\left| l \right\rangle$  and  $\left| r \right\rangle$  of an ordinary (spinless) TS, and  $\alpha$ an angle between the quantization axes  
$z_{\left| l \right\rangle}$ and $z_{\left| r \right\rangle}$ .

The eigenvalues of the Hamiltonian (\ref{eq1}) read:
\begin{eqnarray}\label{eq3}
U_{1, 4} =\mp \frac{1}{2}\sqrt{\left(|u|+\sqrt{h^{2} +\delta ^{2} } \right)^{2} +d^{2} }, \nonumber \\
U_{2, 3} =\mp \frac{1}{2} \sqrt{\left(|u|-\sqrt{h^{2} +\delta ^{2} } \right)^{2} +d^{2} }.
\end{eqnarray}
Previously one of the present authors has shown that under certain temperature and field conditions the PTS model is applicable to a special type of paramagnetic impurity-hole complexes [FeO$_{4}$]$^{0}$, 
presumably present in the Fe$^{3+}$-contaminated silicate glasses \cite{Borisenko}. Under certain model approximations, this resulted into a good agreement with experimental data for the specific heat and dielectric 
susceptibility of several types of iron-contaminated glasses as functions of temperature and applied magnetic field  \cite{Borisenko}.

The idea of existence of the [FeO$_{4}$]$^{0}$ complexes in silicate glasses is fully compatible with the explanation, given by Castner et al.  \cite{Castner} for the broad EPR absorption line at  $g\approx 6$ , detected 
already in the pioneering work by Sands  \cite{Sands}. It is also supported by the evidence for existence of the, to them closely related, [FeO$_{4}$]$^{-}$ complexes, coming from interpretation of EPR spectra in quartz \cite{Mombourquette}. 

\begin{figure}
\includegraphics{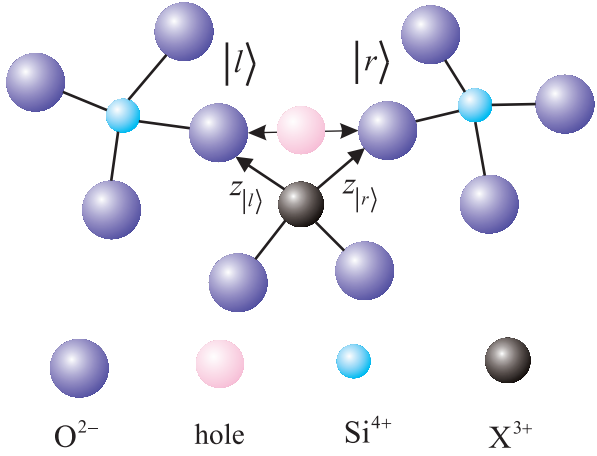}
\caption{\label{Fig2} (Color online) 2D schematic structure of the impurity-hole complex [XO$_{4}$]$^{0}$.}
\end{figure}

In the present paper we consider the paramagnetic impurity-hole complexes of the general type [XO$_{4}$]$^{0}$, where X is a 3-valent substitution paramagnetic impurity. A schematic structure of the [XO$_{4}$]$^{0}$ 
complex is shown in Fig.\ref{Fig2}. 
The ion X$^{3+}$ substitutes Si$^{4+}$ in the silica elementary cell [SiO$_{4}$]. Then one of the neighboring oxygen ions traps a hole to become O$^{-}$ (meaning the O$^{2-}$+hole complex) and to compensate the 
local charge defect  \cite{Schirmer}. This structure is isomorphic to the well-known [AlO$_{4}$]$^{0}$ color center in $\alpha$-quartz \cite{Nuttall&Weil}. In the tetrahedral cage of oxygen ions, with the hole localized at one of them, the ion 
X$^{3+}$ is subject to a crystal field with an approximate 3-fold axial symmetry along $z$, this being the local basis axis, directed from X$^{3+}$ towards O$^{-}$. A gradient of this crystal field couples to the 
electronic quadrupole moment of X$^{3+}$ and causes the quadrupole splitting of its spin states. The spin $J$ of the paramagnetic ion also couples to the spin  $j=1/2$  of the hole and to the external magnetic field $\bf B$. 
Other interactions, such as the nuclear quadrupole, dipole electron-nuclear and nuclear Zeeman ones, are much smaller and may be neglected. 
Then, neglecting a possible nonaxiality of the local crystal field and the dipole 
interaction, the spin Hamiltonian of the [XO$_{4}$]$^{0}$ complex in either coordinate state may be considered in the following form:
\begin{equation}\label{eq4}
\hat{H}_{\rm{spin}} =\beta \left(g_{\rm{O}} \hat{\mathbf {j}}+g_{\rm{X}} \hat{\mathbf{J}} \right)\cdot \mathbf{B} +  D_{z} \hat{j} _{z} \hat{J} _{z} + Q_{z} \hat{J} _{z}^{2}.
\end{equation}
Here $\beta$ is the Bohr magneton, $g_{\rm{O}} \approx 2$ and $g_{\rm{X}}$ are the Land\'e factors for the ions O$^{-}$ and X$^{3+}$, respectively, and  $D_{z} <0$ and $Q_{z} <0$ are the dipole and quadrupole interaction constants, 
respectively. 

In the experimentally relevant range of magnetic fields $B\lesssim J\left|D_{z} \right|/g_{\rm{O}} \beta \cos \left(\alpha /2\right) \sim0.1...1$~T, the ground state of the Hamiltonian (\ref{eq4})  is a quadruplet of spin states $\left| \pm J, \pm 1/2 \right\rangle$. 
The first (non-vanishing) order corrections from the Zeeman and dipole interactions to the ground-state energy level $Q_{z} J^{2}$ are as follows:
\begin{align}\label{eq5}
&E_{1, 3}^{(1)} =-Jg_{X} \beta B_{z} \mp \nonumber \\ 
&\tfrac{1}{2} \sqrt{g_{O}^{2} \beta ^{2} \left(B_{x}^{2}+B_{y}^{2} \right)+\left(JD_{z}-g_{O} \beta B_{z} \right)^{2}}, \nonumber \\ 
&E_{2, 4}^{(1)} =Jg_{X} \beta B_{z} \mp \nonumber \\ 
&\tfrac{1}{2} \sqrt{g_{O}^{2} \beta ^{2} \left(B_{x}^{2}+B_{y}^{2} \right)+\left(JD_{z}+g_{O} \beta B_{z} \right)^{2}}.
\end{align}
In the temperature range $T\lesssim \left|D_{z} \right|/k_{\rm{B}} \sim 0.1$~K the doublet of levels 1 and 2 
is the only thermodynamically relevant one. In this case the Zeeman splitting value is \cite{bisector}:
\begin{equation}\label{eq6}
u\approx 2\beta \left(\tfrac{1}{2} g_{\rm{O}}+Jg_{\rm{X}} \right)B\cos \left(\alpha /2\right).
\end{equation}
In the temperature range $\left|D_{z} \right|/k_{\rm{B}} \lesssim T \lesssim \left|Q_{z} \right|/k_{\rm{B}} \sim 10$~K the whole quadruplet of 
levels (\ref{eq5}) is thermodynamically relevant. Strictly speaking, the PTS model in its simplest form (\ref{eq1}) is inapplicable to the case of more than two relevant 
spin states. Nevertheless, one can hope that it gives a qualitatively adequate physical picture even in this case. 
At  $B=0$  the only gap in the spectrum 
(\ref{eq5}) 
is due to the dipole interaction: $u\left(0\right)=J\left|D_{z} \right|$. 
For the purpose of illustration of the model's results at finite 
values of $B$, as an heuristic assumption, we take the gap between the outside 
levels 1 and 4 as the Zeeman splitting value \cite{bisector}:
\begin{equation}\label{eq7}
u\approx J\left|D_{z} \right|+2\beta g_{\rm{X}} JB\cos \left(\alpha /2\right).
\end{equation}
%
%
%

Now, if one considers tunnel transitions between the  states   $\left| l \right\rangle$ and $\left| r \right\rangle$ in Fig.\ref{Fig2}, the PTS model becomes applicable to this physical object.

In this paper we study the PTS contribution  to the amplitude of the two-pulse (spontaneous) polarization echo and compare the results to experimental data for several glasses with Fe$^{3+}$, Cr$^{3+}$ and Nd$^{3+}$ paramagnetic impurities. 

The essence of the spontaneous polarization echo phenomenon in glasses is as follows. At  $t=0$  the first microwave pulse with frequency 
$\omega _{0}$ and duration $\tau _{1}$ creates a macroscopic coherent superposition of a subsystem of TS with energy gaps $\Delta U\approx \hbar \omega _{0}$. 
This macroscopic coherence quickly vanishes due to the wide distribution of the local TS parameters, but the microscopic (individual phase) coherence 
remains much longer. At $t=t_{w}-\tau _{2}/2$ the second pulse with duration $\tau _{2}$ reverses the temporal evolution of the TS. Then at 
$t\approx 2t_{w}$ the macroscopic coherence is spontaneously restored in the form of an echo.

An external harmonic electric field with an amplitude $\mathbf{F}_{0}$ couples to the PTS dipole momentum operator 
$\hat{\mathbf{P}}=\tfrac{1}{2}\mathbf{p}_{0}\hat{1}_{\tau}\otimes\hat{\sigma}_{z}$ and adds 
a perturbation term  $\hat{V}\left(t\right)=\mathbf{F}_{0}\cdot \hat{\mathbf{P}}e^{-i\omega _{0} t} +(\mathbf{F}_{0}\cdot \hat{\mathbf{P}})^{+} e^{i\omega _{0} t}$  into the PTS 
Hamiltonian (\ref{eq1}). In this case one can consider the following form of the PTS wave function:
\begin{equation}\label{eq8}
\Psi _{\rm{PTS}} \left(t\right)=\sum \limits _{n=1}^{4}a_{n} \left(t\right)\exp \left(-iU_{n} t/\hbar \right)  \left| \varphi _{n}  \right\rangle,
\end{equation}
where $\left| \varphi _{n}  \right\rangle$ are the eigenstates, corresponding to the eigenvalues $U_{n}$ (\ref{eq3})  of the PTS Hamiltonian (\ref{eq1}). 
The coefficients $a_{n} \left(t\right)$ obey the normalization condition 
$\sum \limits _{n=1}^{4}\left|a_{n} \left(t\right)\right|^{2}=1$.

If only one gap in the PTS energy spectrum (\ref{eq3}) is close to resonance:
\begin{equation}\label{eq9}
\omega _{mk}=\left(U_{m} -U_{k} \right)/\hbar=\omega _{0} +\varepsilon _{mk}, \quad \left|\varepsilon _{mk} \right|\ll\omega _{0},
\end{equation}
the corresponding coefficients $a_{m}$ and $a_{k}$ are governed by the following equations  \cite{Landau}:
\begin{equation}\label{eq10}
\begin{cases}
ida_{m}/dt=\eta _{mk} \exp \left(i\varepsilon _{mk} t\right)a_{k} \\ 
ida_{k}/dt=\eta _{mk}^{*} \exp \left(-i\varepsilon _{mk} t\right)a_{m},
\end{cases}
\end{equation}
where  $\eta _{mk} =\left\langle\varphi _{m}\right| \mathbf{F}_{0}\cdot \hat{\mathbf{P}}\left| \varphi _{k}\right\rangle/\hbar$ is nonzero only in presence of the  microwave field.

For the purpose of solution of our problem it is convenient to change to new variables:
\begin{equation}\label{eq_b}
b_{n} \left(t\right)=a_{n} \left(t\right)\exp \left(-iU_{n} t/\hbar \right).
\end{equation}
In these notations the system of Eqs. (\ref{eq10}) reads:
\begin{equation}\label{eq10b}
\begin{cases}
idb_{m}/dt=\eta _{mk} \exp \left(i\varepsilon _{mk} t\right)b_{k}+U_{m}b_{m}/\hbar \\ 
idb_{k}/dt=\eta _{mk}^{*} \exp \left(-i\varepsilon _{mk} t\right)b_{m}+U_{k}b_{k}/\hbar.
\end{cases}
\end{equation}
At $t=\tau_{1}$ (immediately after the first pulse) the solution of Eqs. (\ref{eq10b}) is as follows:
\begin{widetext}
\begin{align} \label{tau1}
&b_{k}(\tau_{1})= \frac{ {\Omega_{mk} b_{k}(0)\cos (\Omega_{mk} \tau_{1})+i\sin (\Omega_{mk} \tau_{1})[(\varepsilon_{mk}/2) b_{k}(0)-\eta_{mk}^{*}b_{m}(0)] }e^{-i\varepsilon_{mk} \tau_{1}/2} }{\Omega_{mk} \exp [i U_{k}\tau_{1}/\hbar]}, \nonumber \\
&b_{m}(\tau_{1})= \frac{ {\Omega_{mk} b_{m}(0)\cos (\Omega_{mk} \tau_{1})-i\sin (\Omega_{mk} \tau_{1})[(\varepsilon_{mk}/2) b_{m}(0)+\eta_{mk} b_{k}(0)] }e^{i\varepsilon_{mk} \tau_{1}/2} }{\Omega_{mk} \exp [i U_{m}\tau_{1}/\hbar]}.
\end{align}
\end{widetext}
Here $\Omega _{mk} =\sqrt{\left|\eta _{mk} \right|^{2}+\varepsilon _{mk}^{2}/4}$ is a generalized Rabi frequency.

During the first period of free evolution $\tau_{1}<t \leq t_{w}-\tau_{2}/2$ the non-diagonal coefficients $\eta _{mk}$ are zero, therefore just before the second pulse the solution of Eqs. (\ref{eq10b}) reads:
\begin{align} \label{tw}
&b_{k}(t_{w}-\tau_{2}/2)= b_{k}(\tau_{1})\exp [-i U_{k}(t_{w}-\tau_{2}/2-\tau_{1})/\hbar], \nonumber \\
&b_{m}(t_{w}-\tau_{2}/2)= b_{m}(\tau_{1})\exp [-i U_{m}(t_{w}-\tau_{2}/2-\tau_{1})/\hbar].
\end{align}
Bearing in mind, that only those PTS make a contribution to the echo signal, which undergo a transition (emit or absorb a photon) during the second pulse, one can readily find the solution of Eqs. (\ref{eq10b}) at $t>t_{w}+\tau_{2}/2$ (in the period of free evolution after the second pulse):
\begin{align} \label{t}
&b_{k}(t-t_{w}-\tau_{2}/2)= \nonumber \\
&\frac{ -i\eta_{mk}^{*}\sin (\Omega_{mk} \tau_{2})b_{m}(t_{w}-\tau_{2}/2)e^{-i\varepsilon_{mk} \tau_{2}/2} }{\Omega_{mk} \exp [i U_{k}(t-t_{w}+\tau_{2}/2)/\hbar]}, \nonumber \\
&b_{m}(t-t_{w}-\tau_{2}/2)= \nonumber \\
&\frac{ -i\eta_{mk}\sin (\Omega_{mk} \tau_{2})b_{k}(t_{w}-\tau_{2}/2)e^{i\varepsilon_{mk} \tau_{2}/2} }{\Omega_{mk} \exp [i U_{m}(t-t_{w}+\tau_{2}/2)/\hbar]}.
\end{align}
The solution of Eqs. (\ref{eq10b}) in the form Eqs. (\ref{tau1}) - (\ref{t}), with the initial conditions 
\begin{align} \label{eq11}
&\left|b_{i} \left(0\right)\right|^{2}=\left|a_{i} \left(0\right)\right|^{2} =n_{i}^{\rm{eq}} \left(T\right)=\nonumber \\ 
&\exp\left(- \frac{U_{i}}{ k_{\rm{B}} T}\right) \Bigg/ \sum \limits_{j=1}^{4}\exp \left(- \frac{U_{j}}{ k_{\rm{B}}T} \right),
\end{align}
given by the Boltzmann population of the PTS levels (\ref{eq3})  before the first pulse, can be used to find the time dependence of the PTS polarization:
\begin{equation}\label{eq12}
A_{\rm{PTS}}\left(t\right)=\left\langle\Psi _{\rm{PTS}}^{*}\left(t\right)\right|\mathbf{F}_{0}\cdot \hat{\mathbf{P}}\left|\Psi _{\rm{PTS}}\left(t\right)\right\rangle/F_{0}.
\end{equation}
Neglecting all possible dynamic-dephasing mechanisms such as the interaction with phonons and with neighboring TS and omitting the terms odd in $\eta_{mk}$ (they vanish due to the isotropic angular distribution of the PTS dipoles $\mathbf{p}_{0}$),
 one can derive from Eq. (\ref{eq12}) an explicit formula for the echo amplitude (envelope):
\begin{equation}\label{eq13}
A_{\rm{PTS}}\left(t'\right)=A_{0} \sum \limits _{m>k}|\eta_{mk}| \left(n_{k}^{\rm{eq}}-n_{m}^{\rm{eq}}\right) H_{mk} \left(t'\right),
\end{equation}
where
\begin{align}\label{eq14}
&H_{mk} \left(t'\right)=\sin ^{2} \left(\Omega _{mk} \tau _{2} \right)\Big[\sin \left(2\Omega _{mk} \tau _{1} \right) \cos \varepsilon _{mk} t'+ \nonumber \\
&\left(\varepsilon _{mk}/\Omega _{mk}\right) \sin ^{2} \left(\Omega _{mk} \tau _{1} \right)\sin \varepsilon _{mk} t'\Big]\left|\eta _{mk} \right|^{3}\Big/\Omega _{mk}^{3}.
\end{align}
Here $A_{0}\propto\hbar/F_{0}$ is a prefactor and $t'=t-2t_{w}+\tau _{1}$. In Eq. (\ref{eq14}) we have omitted the terms odd in  $\varepsilon$, since they vanish after integration over this variable (see Eq. (\ref{eq16}) below).

An average echo amplitude is a result of integrating Eq. (\ref{eq13}) with a distribution function for the double-well potential parameters $f\left(h, \Delta , \alpha \right)$. In practice, it is convenient to change one of the 
integration variables to $\varepsilon$, e.g. $h\rightarrow\varepsilon$. As an approximation, we integrate only the "fast" function of  $\varepsilon$  Eq. (\ref{eq14}). The remaining ones are taken at the resonance hypersurface:
\begin{equation}\label{eq15}
\varepsilon _{mk} \left(h, \Delta , \alpha, u \right)=0.
\end{equation}
This is valid since Eq. (\ref{eq14}) is nonzero in the narrow region $\left|\varepsilon _{mk} \right| \lesssim \left|\eta _{mk} \right|\propto\left|\mathbf{p}_{0}\cdot \mathbf{F}_{0}\right|/\hbar \ll \omega _{0}$. In this way one is able to 
represent the average echo amplitude as follows:
\begin{widetext}
\begin{equation}\label{eq16}
\bar{A}_{\rm{PTS}} \left(t'\right)=A_{0} \int \limits _{0}^{\Delta _{\rm{max}}}d\Delta \int \limits _{0}^{2\pi}d\alpha \sum \limits _{m>k}\sum \limits _{h_{mk} \left(\Delta, \alpha, u \right)}\left\{|\eta_{mk}|  \left(n_{k}^{\rm{eq}}-n_{m}^{\rm{eq}} \right)\left(\frac{\partial \varepsilon _{mk}}{\partial h}\right)^{-1} f\left(h, \Delta , \alpha \right)\int \limits _{-\infty }^{\infty}d\varepsilon _{mk} H_{mk} \left(t'\right) \right\},
\end{equation}
\end{widetext}
where $h_{mk} \left(\Delta , \alpha, u \right)$ are the real roots of Eq. (\ref{eq15}).

An integrated echo amplitude is a result of integrating Eq. (\ref{eq16}) over  $t'$:
\begin{equation}\label{eq17}
\bar{A}_{\rm{int}} =\int \limits _{-\infty }^{\infty }dt'\bar{A}_{\rm{PTS}} \left(t'\right).
\end{equation}
To compare the results of the PTS model to the available experimental data we use the following model distribution function for the double-well potential parameters:
\begin{align}\label{eq18}
&f\left(h, \Delta , \alpha \right)=\frac{1}{\sqrt{2\pi}\delta h} \exp \left(-\frac{h^{2}}{2 \delta h^{2}} \right) \times \nonumber \\
&\frac{1}{\Delta} \exp \left(-\frac{\Delta _{0}}{\Delta} \right)\cdot \frac{1}{\sqrt{2\pi} \delta \alpha} \exp \left[-\frac{\left(\alpha -\alpha _{0} \right)^{2}}{2 \delta \alpha ^{2}} \right],
\end{align}
with  $\delta h/k_{\rm{B}}=230$~K, $\Delta _{0}/k_{\rm{B}}=40$~mK, $\Delta _{\rm{max}}/k_{\rm{B}}=4$~K and the parameters $\alpha _{0}$ and $\delta \alpha$ depending on the substitution impurity as listed in Table~\ref{Table2}.
Eq. (\ref{eq18}) is an analytic function of $\Delta$: it is zero at $\Delta=0$, then passes through a maximum at $\Delta=\Delta_{0}$ and falls asymptotically as $1/\Delta$ at $\Delta \gg \Delta_{0}$, similar to the standard TS distribution function (see e.g. \cite{TM}).
\begin{table*}
\caption{\label{Table2} Fitting parameters for the different [XO$_{4}$]$^{0}$ complexes.} 
\begin{ruledtabular}
\begin{tabular}{cccccccc}
\textrm{Complex} & $J$ & $g_{X}$ & $\alpha_{0}$ & $\delta \alpha$ & $J\left|D_{z} \right|/k_{\rm{B}}$, \textrm{(K)} & $\left|\mathbf{p}_{0}\cdot \mathbf{F}_{0}\right|\tau _{1}/\hbar$ & $\left|\mathbf{p}_{0}\cdot \mathbf{F}_{0}\right|\tau _{2}/\hbar$ \\
\hline
 $\rm{[FeO}_{4}\rm{]}^{0}$  & 5/2 & 2 & 2.6  &  0.1  &  -  &  0.819  &  1.638 \\ 
$\rm{[CrO}_{4}\rm{]}^{0}$  &  3/2\footnote{These values correspond to the zero orbital momentum of X$^{3+}$ and, therefore, to the minimal ionic radius to minimize the positive size misfit of X$^{3+}$ compared to Si$^{4+}$.} & 2\footnotemark[1] & 2.07  &  0  &  0.354  &  2.235  &  2.235  \\
$\rm{[NdO}_{4}\rm{]}^{0}$  &  3/2\footnotemark[1] & 2\footnotemark[1] & 2.42  &  0.1  &  0.22  &  1.117  &  1.117 \\
\end{tabular} 
\end{ruledtabular} 
\end{table*}
%

As it was stated above, the temporal attenuation of the echo signal is out of the scope of our consideration. Ludwig et al.  \cite{Ludwig_PRL} found experimentally that in the barium-aluminosilicate glass (BAS), studied at 1~GHz 
and 10~mK, the integrated echo amplitude depends on the waiting time exponentially: $A_{\rm{int}} \propto \exp \left(-t_{w}/\tau\right)$, with the relaxation time  
$\tau$ being practically independent on the value of applied magnetic field. Therefore, one can conclude that the relative value of the echo amplitude, 
e.g. $A\left(B\right)/A\left(0\right)$, should be independent (roughly, at least) on the experimental value of the waiting time $t_{w}$. In this way one also gets rid of the main temperature dependence of the 
echo signal due to the temperature variation of the relaxation time $\tau$, with the only residual temperature dependence coming from the initial equilibrium levels' population  $n_{i}^{\rm{eq}}\left(T\right)$, given by Eq. (\ref{eq11}).

From the numerical studies of Eq. (\ref{eq16}) as a function of applied magnetic field at $t'=0$ we find that it reaches a minimum at some value $B_{\rm{min}}$ and that within our accuracy $\bar{A}_{\rm{PTS}} \left(B_{\rm{min}}\right)\approx 0$. 
In terms of the Zeeman splitting $u$ the minimum occurs at $u_{\rm{min}}\approx \hbar\omega_{0}$.

\begin{figure}
\includegraphics{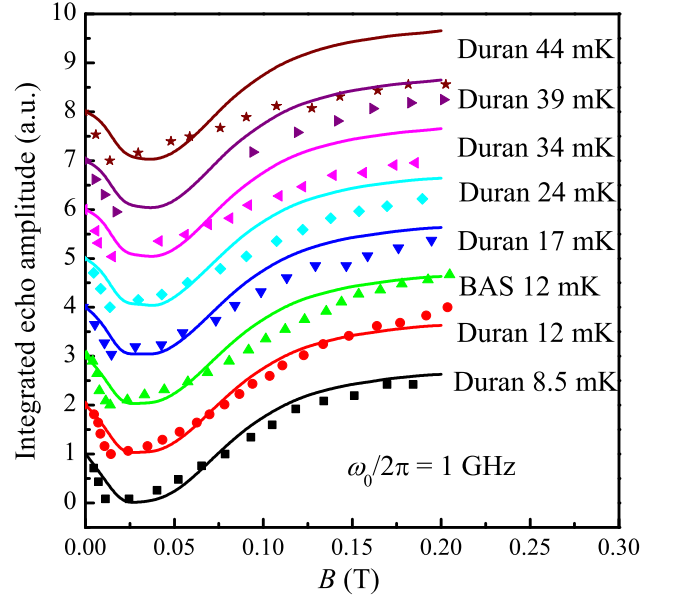}
\caption{\label{Fig3} (Color online) Integrated echo amplitude vs. $B$. The points stand for experimental data \cite{Ludwig_JLTP}. The curves are calculated from the present theory. The data are shifted arbitrarily along the vertical axis. Details as given in the text.}
\end{figure}

With the above speculations in mind, in Fig.~\ref{Fig3} we re-plot experimental data  \cite{Ludwig_JLTP} for the integrated echo amplitude in the Fe-contaminated glasses Duran and BAS in the form 
$\left[A_{\rm{int}}\left(B\right)-A_{\rm{int}}\left(B_{\rm{min}}\right)\right]/\left[A_{\rm{int}}\left(0\right)-A_{\rm{int}}\left(B_{\rm{min}}\right)\right]$ (to subtract the magnetic-independent contribution, coming from the ordinary TS) 
and compare them to the function $\bar{A}_{\rm{int}}\left(B\right)/\bar{A}_{\rm{int}}\left(0\right)$, calculated from Eq. (\ref{eq17}), with the Zeeman splitting calculated from Eq. (\ref{eq6}) and with the parameters, 
corresponding to the [FeO$_{4}$]$^{0}$ complex, as listed in Table~\ref{Table2}. 
%
%

\begin{figure}
\includegraphics{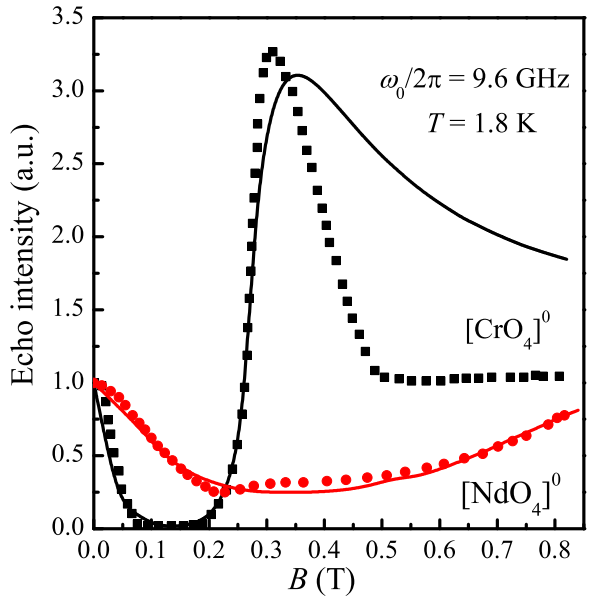}
\caption{\label{Fig4} (Color online) Intensity of the echo signal vs. $B$. The points stand for experimental data \cite{Smolyakov&Solovarov}. The curves are calculated from the present theory. Details as given in the text.}
\end{figure}

In Fig.~\ref{Fig4} we re-plot experimental data  \cite{Smolyakov&Solovarov} for the echo intensity in the Cr- and Nd-contaminated glasses in the form  $I\left(B\right)/I\left(0\right)$ together with the function 
$\bar{A}_{\rm{PTS}}^{2}\left(B\right)/\bar{A}_{\rm{PTS}}^{2}\left(0\right)$, calculated from Eq. (\ref{eq16}) at $t'=0$, with the Zeeman splitting calculated from Eq. (\ref{eq7}) and with the parameters for the corresponding complexes as listed in Table~\ref{Table2}. 
In the case of the Nd-contaminated glass, for the purpose of fitting, we added a constant to Eq. (\ref{eq16}), which accounts for the magnetic-independent contribution, presumably coming from the non-magnetic TS, 
probably of the [AlO$_{4}$]$^{0}$ or [BO$_{4}$]$^{0}$ types.
%

Yet another interesting phenomenon, reported \cite{Smolyakov&Solovarov} for the Cr-contaminated glass, is a dynamic enhancement of the echo intensity during a rapid decrease of the magnetic field from a constant value in the vicinity of 
$B_{max} \approx 0.3$~T, corresponding to the maximum of the echo intensity (see Fig.\ref{Fig4}). In the framework of the present theory we provide the next qualitative explanation for this phenomenon. 
Since the decrease of the magnetic field leads to the decrease of the Zeeman splitting $u$ (see Eq. (\ref{eq7})), the population of the PTS levels turns into a non-equilibrium one, namely, the excited levels become under-populated. 
The equilibration process goes through the absorption of phonons by the PTS. This "pumping" of the energy from the phonon subsystem into the PTS one leads to the decrease of the sample temperature $T$. 
This process is analogous to the well-known adiabatic demagnetization cooling of paramagnetic samples. Then, the decrease of the sample temperature $T$ leads to the slowing down of the phonon attenuation of the echo signal and, 
therefore, to its enhancement.

It should be mentioned that, from the studies of the dielectric relaxation phenomena in smoky quartz \cite{Vos&Volger}, containing the [AlO$_{4}$]$^{0}$ complexes, one can expect the characteristic relaxation rate due to the interaction with 
phonons for the [XO$_{4}$]$^{0}$ complexes to be much slower than for the ordinary TS. That is why in the echo experiments they manifest themselves even in the liquid helium temperature range, where the contribution from the 
ordinary TS is already damped by phonons.

It seems reasonable that the echo response in the glasses free from paramagnetic impurities, such as the borosilicate glass BK7 \cite{Ludwig_JLTP} and deuterated glycerol \cite{Glycerol}, can be explained in the framework of the 
quadrupole model \cite{Quadrupole}. 
The limited applicability of the quadrupole model explains its success in reproducing the experimental data only for BK7 \cite{Quadrupole,Shumilin&Parshin} and glycerol \cite{Shumilin&Parshin}.

It is tempting to assume, that TS of the same type, but originating from the non-magnetic substitution impurities, such as the [AlO$_{4}$]$^{0}$ or [BO$_{4}$]$^{0}$ ones, also give their contribution to the low-temperature properties of 
multisilicate glasses and may be responsible for some deviations of experimental data from the predictions of the standard tunneling model.

In summary, we have demonstrated, that with the assumptions made, the idea of application of the PTS model to the [XO$_{4}$]$^{0}$ paramagnetic impurity-hole complexes  results into a systematic reasonable 
agreement with the experimental data from the spontaneous polarization echo in different silicate glasses with different paramagnetic impurities and in a wide range of frequencies and temperatures.

\end{document}